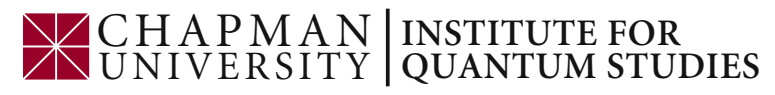

REGULAR PAPER

# On the applicability of Kolmogorov's theory of probability to the description of quantum phenomena. Part I: foundations

**Maik Reddiger**

**Abstract** By formulating the axioms of quantum mechanics, von Neumann also laid the foundations of a "quantum probability theory". As such, it is regarded a generalization of the "classical probability theory" due to Kolmogorov. Outside of quantum physics, however, Kolmogorov's axioms enjoy universal applicability. This raises the question of whether quantum physics indeed requires such a generalization of our conception of probability or if von Neumann's axiomatization of quantum mechanics was contingent on the absence of a general theory of probability in the 1920s. This work argues in favor of the latter position. In particular, it shows how to construct a mathematically rigorous theory for non-relativistic $N$-body quantum systems subject to a time-independent scalar potential, which is based on Kolmogorov's axioms and physically natural random variables. Though this theory is provably distinct from its quantum-mechanical analog, it nonetheless reproduces central predictions of the latter. Further work may make an empirical comparison possible. Moreover, the approach can in principle be adapted to other classes of quantum-mechanical models. Part II of this series discusses the empirical violation of Bell inequalities in the context of this approach. Part III addresses the projection postulate and the question of measurement.

M. Reddiger
Department of Computer Science and Languages, Anhalt University of Applied Sciences, Lohmannstraße 23, 06366 Köthen (Anhalt), Germany

M. Reddiger (✉)
Center for Research, Transfer, and Entrepreneurship, Anhalt University of Applied Sciences, Hubertus 1a, 06366 Köthen (Anhalt), Germany
e-mail: maik.reddiger@hs-anhalt.de

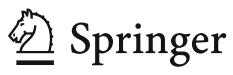

## 1 Introduction

For almost a century now, the theory of quantum mechanics has stood firm at the core of our modern understanding of the physical microcosm. Still, as evinced by the multitude of interpretations in existence today [1], the theory continues to mystify even experts in the subject. This state of affairs has given quantum mechanics the idiosyncratic status of what one could call a well-established theory with contested foundations [2,3].

So far, the following aspect of this problem has only received comparatively little attention [4–16]: across the empirical sciences, quantum mechanics and its descendants are the only scientific theories that are not based on the mathematical theory of probability, as developed by Kolmogorov in 1933 [17,18]. Since von Neumann's axiomatization of quantum mechanics [19–23] predates Kolmogorov's work, it therefore stands to reason, whether quantum physics indeed requires a different notion of probability or if von Neumann's axiomatization of quantum mechanics was contingent on the absence of a general, mathematical theory of probability in the 1920s.

Should Kolmogorov's theory of probability indeed be applicable to the description of quantum phenomena, then "quantum probability theory" would lose its physical support and contemporary quantum theory, both non-relativistic and relativistic, would require a thorough revision in terms of its underlying probability theory.

While the need for a "quantum probability theory" is widely defended in the literature [5,6,10–12], the counter-hypothesis has, so far, not been ruled out with certainty. The mere incompatibility of quantum mechanics with "classical probability theory", as various so-called "no go theorems" [22–26] aim to show, can hardly be taken serious as physical arguments (cf. [27] and Chap. 1 in [14]), for those derive their relevance from the following implicit assertion: quantum mechanics is the only (non-relativistic) theory, that can correctly describe the data. On the contrary, it is an epistemological truism, that no amount of data can uniquely determine a single scientific theory. The scientific method demands that competing theories are empirically tested and compared on their own merits, if possible. That research in the history of the subject suggests the presence of a long-standing, systematic bias against theoretical approaches competing with (Copenhagen) quantum mechanics [2,28,29], emphasizes the need for both scientific rigor and openness in this debate.

Accordingly, the overarching physical question of the applicability of Kolmogorov's theory of probability to this domain of science can only be indirectly addressed by constructing specific competitor theories on this basis and following the scientific method.

This work shows that for non-relativistic $N$-body quantum systems subject to a time-independent scalar potential, it is indeed possible to construct a theory based on Kolmogorov's axioms and physically natural random variables, which reproduces central predictions of quantum mechanics (cf. Lemma 1, Theorem 1, and Proposition 1). Furthermore, the constructed theory is shown to be capable of making distinct predictions (cf. Example 3), so that the two theories can in principle be compared empirically.

The main line of reasoning is as follows: first and with a historical focus, we consider the general relationship between mathematical probability theory and quantum mechanics in Sect. 2. Section 3 provides a probabilistic view on the double slit experiment and quantum theory as a whole, in line with the *ensemble interpretation* of quantum mechanics [30,31]. This physically motivates the implementation of Kolmogorov's axioms in quantum theory via the Born rule for position (Sect. 4, in particular Lemma 1). Corresponding observables can be obtained from Madelung's reformulation of the Schrödinger equation (Sect. 5), in line with [13]. This suggests that the theory, *geometric quantum theory* [13], can be erected upon the Madelung equations in conjunction with the Born rule. Yet, a recent review [15] showed that the mathematical theory of the Madelung equations is still underdeveloped. Hence and at least for the time being, another approach is needed. Instead, one may develop a "hybrid theory", in which the dynamics is borrowed from quantum mechanics and the basic quantities are rigorously defined in the context of said dynamics. For the aforementioned class of quantum systems, this is done in Sect. 6. In result, an alternative theory for the aforementioned quantum system, which is based on Kolmogorov's axioms, is indeed obtained.

The reader, who is primarily interested in the analytical results of this work, is invited to skip Sects. 2, 3, and 5, consulting Table 1 when needed. A synopsis of the theory developed here and a guideline for how to construct similar ones from other (classes of) quantum-mechanical models is given in Sect. 7.

**Table 1** Direct comparison of the two probability theories

| | von Neumann (pure state) | Kolmogorov |
|---|---|---|
| Mathematical setting | Separable, complex Hilbert space $(\mathcal{H}, \langle \cdot, \cdot \rangle)$ | Measure space $(\Lambda, \mathcal{A})$ |
| Object that determines probabilities | State $\Psi \in \mathcal{H}$ with $\sqrt{\langle \Psi, \Psi \rangle} = \|\Psi\| = 1$ | Measure $\mathbb{P}\colon \mathcal{A} \to [0, 1]$ , $A \mapsto \mathbb{P}(A)$ with $\mathbb{P}(\Lambda) = 1$ |
| "Observable"/ random variable | Unbounded self-adjoint operator $\hat{X}\colon \operatorname{dom} \hat{X} \to \mathcal{H}$ with dense domain $\operatorname{dom} \hat{X}$ in $\mathcal{H}$ | Measurable function $X\colon \Lambda \to \mathbb{R}$ |
| Probability to obtain values in $I \in \mathcal{B}(\mathbb{R})$ | $\frac{\lvert\langle \Psi, P_I \Psi \rangle\rvert^2}{\|P_I \Psi\|^2} = \int_I \mathrm{d}\langle \Psi, P_{\cdot} \Psi \rangle$ | $\int_{X^{-1}(I)} \mathrm{d}\mathbb{P} = \int_I \mathrm{d}\big(\mathbb{P} \circ X^{-1}\big)$ |
| Expectation value | $\langle \Psi, \hat{X} \Psi \rangle = \int_{\mathbb{R}} \xi \ \mathrm{d}\langle \Psi, P_{\cdot} \Psi \rangle(\xi)$ for $\Psi \in \operatorname{dom} \hat{X}$ | $\mathbb{E}(X) = \int_\Lambda X \,\mathrm{d}\mathbb{P}$ $= \int_{\mathbb{R}} \xi \,\mathrm{d}\big(\mathbb{P} \circ X^{-1}\big)(\xi)$ for integrable $X$ |

$\mathcal{B}(\mathbb{R})$ denotes the Borel sets on $\mathbb{R}$. The map $I \mapsto P_I$ is the projection valued measure for the operator $\hat{X}$ and, accordingly, $I \mapsto \langle \Psi, P_I \Psi \rangle$ is the spectral measure (cf. Sec. VIII.3 in [32]). Note the close analogy between the spectral measure and $\mathbb{P} \circ X^{-1}$. For an introduction to mathematical probability theory, the reader is referred to [33]. Operator theory with a focus on quantum mechanics is treated, for instance, in [34]

## 2 Probability and quantum mechanics

The two modern mathematical theories of probability were developed by the Hungarian mathematician John von Neumann (1903–1957) [19–23] and the Russian mathematician Andrej N. Kolmogorov (1903–1987) [17,18]. That today, there are two such theories, points at a potential discrepancy in the modern foundations of probability theory (as a structural theory of nature).

Historically, it is known that both researchers knew each other and followed each other's work [35]. Still, the author did not find any documented evidence of a discussion among the two about this particular topic, even after searching in the literature and a consultation of a monograph on the history of modern probability theory [36]. On the contrary, Parthasarathy [37] writes that Kolmogorov showed little interest in quantum mechanics, adding that "it is a blessing to have enough ground left for future mathematicians to till."

To the author's knowledge, the Hungarian physicist Fényes was the first person in the literature to claim the validity of Kolmogorov's axioms in the quantum domain [7]. The physical need for a "quantum probability theory" was explicitly questioned by Szabó [4], though on the basis of different arguments and following a different approach than the one pursued here.

A direct comparison of the two probability theories for pure states is given in Table 1. In the contemporary literature, this conflict in the foundations of probability is resolved by showing that von Neumann's "quantum probability theory" may be viewed as a "non-commutative" generalization of the "classical probability theory" discovered by Kolmogorov [4,5,10,11].

While this accords with the general philosophy that quantum mechanics is to be viewed as a generalization of "classical mechanics", the particular historical context of the development of the two theories may be put forward as an argument against this conventional point of view.

Mostly guided by the ideas of Heisenberg, Dirac, Born, and Jordan, von Neumann developed his theory of probability with the explicit goal of axiomatizing quantum mechanics [19,38]. That, and not the development of a probability theory, was his aim. Indeed, in his most important article on the topic [19] von Neumann noted that he was merely formalizing "the common calculus of probability." Clearly, he was not aware that one day, his notion of probability would be in conflict with what is generally associated with that term.

Kolmogorov, on the other hand, took a direct approach to the question via measure theory. His fundamental work [17,18] was published in 1933. This is one year after von Neumann's famous book on the mathematical foundations of quantum mechanics [22,23] was published and a total of 6 years after von Neumann's article on the topic [19] appeared.

*The common view in the literature therefore implicitly claims, that von Neumann generalized Kolmogorov's theory before the latter could even publish his approach.*

One should add to this, that today Kolmogorov's theory of probability is successfully employed in virtually all of the sciences, except quantum theory. Convincing examples outside of quantum physics, where von Neumann's probability theory is required, seem to be missing (see [6] for a different view).

Therefore, the hypotheses put forward here are that, one, von Neumann's axiomatization was contingent on the absence of a general, mathematical theory of probability at the time and, two, that the physically correct theory of probability is indeed the one discovered by Kolmogorov.

If this is the case, then contemporary quantum theory, both non-relativistic and relativistic, requires a thorough revision in terms of its underlying probability theory.

To put these hypotheses to scientific scrutiny, the question of whether an empirically successful quantum theory can be constructed on the basis of Kolmogorov's axioms is to be investigated. What follows shows that this may indeed be possible.

## 3 Physical considerations

Before addressing any mathematical details, the reader shall be briefly introduced to the central physical ideas.

For this purpose, we begin with a discussion of the double slit experiment with electrons. The argument in principle also applies to other material particles; on the topic of electromagnetic radiation, the reader is referred to Planck's remark in [39]. The character of the double slit experiment as an "examplar" in the sense of Kuhn [40] was also recognized by Feynman, who wrote that "it has in it the heart of quantum mechanics" and that "[i]n reality, it contains the *only* mystery" [41].

For the sake of argument, we shall assume that in the experiment, only a single electron is released at a time, so that one can view the runs as mutually independent. Then, if an electron impinges on the detector screen, it creates a single, pointlike impact. Yet, as one repeats the experiment a large number of times, an interference pattern begins to emerge on the screen.

In textbook accounts, wave functions are commonly associated with each slit and the pattern is said to emerge from an "interference term", resulting out of the absolute modulus squared of the superposition [41]. However, such accounts are heuristic and somewhat misleading, since in quantum mechanics, the pattern must be viewed as a consequence of the time evolution of a single initial wave function in the presence of the double slit [42].

Still, the argument, that this pattern cannot result from particles in the usual sense of the word [41], needs to be addressed. The argument is that the pattern is not the same as the conjunction of the two patterns one obtains with one of the slits closed. That is, if $\sigma_{\text{double}}$ is the (2-dimensional, time-independent) probability density of the impact locations on the detector screen for the double slit experiment and $\sigma_i$ with $i \in \{1, 2\}$ are the probability densities for the experiment with one of the slits closed, then we have

$$\sigma_{\text{double}} \neq \frac{1}{2}(\sigma_1 + \sigma_2). \tag{3.1}$$

Indeed, if the presence of the other slit did not have an influence on the impact location of an individual electron hitting the screen, then in the above equation equality would hold. In the Copenhagen school, this is interpreted as an exhibition of the wave-like nature of electrons, whereas the pointlike impacts are interpreted as an exhibition of their particle-like nature. Wave particle duality states that neither view provides a full picture, they are complementary.

In the 1960s, Landé expressed a different perspective on the matter: according to him [43], the electrons are pointlike particles and it is only their collective, statistical behavior that is wave-like.

Indeed, the attentive reader may have noticed that the above argument does not rule out that electrons are particles, it only rules out naive notions of locality. For instance, an electron could pass through one of the slits, take a turn into the second slit and then go back through the first one before hitting the screen. Or, following de Broglie [3], one could imagine that the electron is pointlike, yet accompanied by a wave field in its immediate vicinity, which interacts with the other slit and in result changes the electron trajectory [44–46]. Such trajectories must differ, when one of the slits is closed. Hence, both of these behaviors could explain Eq. (3.1), at least qualitatively. Of course, the argument put forward here is that physically peculiar behavior can statistically lead to Eq. (3.1), it is not to commit to a particular explanation. In this sense, it seems difficult to evade the conclusion that Landé's criticism of wave particle duality is indeed justified.

What Landé advocated for is *a probabilistic understanding of quantum theory*: while quantum theory does not predict the positions of the particles, it does predict the respective probabilities. Following Ballentine [30,31], quantum theory is concerned with *ensembles of similarly prepared systems (of particles and radiation)*. An ensemble is a theoretical abstraction, consisting of an infinite number of samples. Each *sample* is a particular quantum system in the aforementioned sense (though we shall not be overly pedantic in our usage of the term "quantum system" here). In line with a frequentist understanding of probability [47] and staying on a heuristic level, the statistics of $n$ samples approach the respective probabilities of the ensemble as $n \to \infty$, as long as the random variables in question are mutually independent and identically distributed for each run.

In the above example, this means that the relative impact frequencies of the electrons hitting the screen after $n$ runs will approach the probability density $\sigma_{\text{double}}$ as $n \to \infty$, appropriately understood. For $\sigma_1$ and $\sigma_2$, this is also the case, yet they refer to different physical contexts, different "similarly prepared systems" that is, and therefore belong to distinct ensembles.

To summarize the argument: the results of the double slit experiment do not conflict with Kolmogorov's theory of probability, at least on a qualitative level. The general claim is that this is also true for other quantum phenomena.

Rather, the applicability of Kolmogorov's axioms to this physical domain is to be tested quantitatively. Again, this can only be achieved indirectly by constructing and testing respective physical theories.

*Remark 1* Historically, there have been many objections to such probabilistic approaches. The interested reader is referred, for instance, to [22–26], as well as Chap. 5 in Moretti's book [34]. Addressing all of these objections individually would go beyond the scope of this article, so we focus on those, which are arguably the strongest in this context.

First, if one follows the common interpretation of the uncertainty principle, then, in quantum theory, the notion of a point particle, let alone its trajectory, is not a sensible concept. The common interpretation is, however, not uncontested. Already in the 1930s, Popper objected to it on the basis of a probabilistic view on quantum theory (see [48], §78 in [49], and [50] for a historical account): the assertion that quantum theory is concerned with probabilities implies that Heisenberg's inequality cannot be interpreted as a statement on an individual particle. On the contrary, it ought to be understood as a probabilistic statement on the respective (quantum-mechanical) standard deviations (for the ensemble). That is, in the ensemble interpretation, Heisenberg's inequality is not a statement on the "localizability" of individual particles and can therefore not be employed as an argument against the particle model. Landé [43] and Fényes [7] argued along similar lines.

Second, the most direct result on the applicability of Kolmogorov's axioms to quantum phenomena, which the author is aware of and which is empirically relevant, is the BCHSH inequality (cf. [34,51] and also Sec. 8 in [52] for a refutation of GHZ-type arguments). This constitutes a generalization of Bell's theorem [53], that overcomes the "unrealistic restriction that for some pair of parameters [...] there is perfect correlation" [51]. The latter can be derived on the basis of Kolmogorov's axioms and the additional assumption of *Bell locality*: each detector random variable $A$ and $B$ only depends on the setting of the respective detector, not on the setting of the other.

Though there are several variations of such Bell inequalities, it is generally acknowledged that they are empirically violated [54–59]. It is noteworthy that such results are necessarily of statistical nature and that statistical means of sample data generated by "local realistic" models need not always satisfy a respective inequality for the means. Still, provided that the trials are mutually independent and the products of $A$ and $B$ are identically distributed for each setting, the central limit theorem makes the empirical investigation of the violation of Bell inequalities feasible. Indeed, some experiments were able to meet the conventional threshold of a violation by 5 standard deviations.

Part II of this series reviews the underlying mathematical argument and discusses the implications of those empirical findings for quantum theories, which are based on mathematical probability theory. ◇

## 4 Born's rule as an approach for basing quantum theory on Kolmogorov's axioms

The Born rule for position provides a natural approach to the construction of a quantum theory on the basis of Kolmogorov's axioms [60]. This section shows how to do this for an $N$-body quantum system "without spin" ($N \in \mathbb{N}$).

We consider a time-dependent, normalized (Schrödinger) wave function $\Psi_t \in L^2(\mathbb{R}^{3N}, \mathbb{C})$ on the configuration space $\mathbb{R}^{3N}$ and denote by $\mathcal{B}^*(\mathbb{R}^{3N})$ the corresponding $\sigma$-algebra of Lebesgue sets. Then, by Born's rule, the probability that the bodies are in a configuration lying in $W \in \mathcal{B}^*(\mathbb{R}^{3N})$ at time $t \in \mathbb{R}$ is

$$\mathbb{P}_t(W) = \int_W |\Psi_t|^2(\vec{r}_1, \ldots, \vec{r}_N)\ \mathrm{d}^{3N} r. \tag{4.1}$$

Indeed, normalization of $\Psi_t$ is equivalent to $\mathbb{P}_t(\mathbb{R}^{3N}) = 1$. Normalization, in turn, follows from unitary evolution: If, initially, $\Psi_0 \in L^2(\mathbb{R}^{3N}, \mathbb{C})$ and $t \mapsto U_t$ is any 1-parameter family of unitary evolution operators, then $\Psi_t = U_t \Psi_0$ is indeed normalized for any $t \in \mathbb{R}$. Accordingly, the following holds.

**Lemma 1** *Let $N \in \mathbb{N}$, $\Psi_0 \in L^2(\mathbb{R}^{3N}, \mathbb{C})$, and for all $t \in \mathbb{R}$, let $U_t$ be a unitary operator on $L^2(\mathbb{R}^{3N}, \mathbb{C})$. Set $\Psi_t \equiv U_t \Psi_0$ and define the measure $\mathbb{P}_t$ on $\mathcal{B}^*(\mathbb{R}^{3N})$ via Eq. (4.1).*

*Then, the tuple $\left(\mathbb{R}^{3N}, \mathcal{B}^*(\mathbb{R}^{3N}), \mathbb{P}_t\right)$ is a probability space for every $t \in \mathbb{R}$.* ◇

From a mathematical point of view, Lemma 1 is all we require to base any such $N$-body quantum theory on the basis of Kolmogorov's axioms. By applying standard probability theory [33], Lemma 1 determines what a random variable is, how their expectation values and distributions are defined, how conditional probabilities are obtained, etc.

To obtain a physical theory based on Kolmogorov's axioms, however, more is required.

The dynamical evolution $t \mapsto U_t$, one can generally carry over from quantum mechanics. The reason is that this is less difficult to test empirically than the underlying probability theory: a physically incorrect evolution tends to predict physically incorrect emission spectra, scattering cross sections etc., so that, in this respect, quantum-mechanical models are comparatively trustworthy.

The actual difficulties concern the observables and the status of the projection postulate. In particular, the '*observables*' ought to be real random variables on $\left(\mathbb{R}^{3N}, \mathcal{B}^*(\mathbb{R}^{3N})\right)$ instead of (unbounded, densely defined) self-adjoint operators on $L^2(\mathbb{R}^{3N}, \mathbb{C})$. In the next section, we motivate such random variables on the basis of physical considerations. In Sect. 6 thereafter, those random variables are defined rigorously. The projection postulate and the question of measurement are addressed via conditional probabilities in Part III of this series.

## 5 Motivation of random variables from Madelung's equations

In 1926, Madelung discovered a reformulation of the 1-body Schrödinger equation, which is reminiscent Newtonian continuum mechanics [8,13,16,61–63]. His reformulation can be generalized to the $N$-body Schrödinger equation, which is what I shall consider here (cf. [64] and §7.1.2 in [65]).

Assuming, for the sake of simplicity, that the time-dependent wave function $\Psi$ is smooth, we set

$$\rho := |\Psi|^2 \tag{5.1}$$

and

$$\vec{v}_a := \frac{\hbar}{m_a} \mathrm{Im}(\nabla_a \Psi / \Psi) \tag{5.2}$$

for any $a \in \{1, \ldots, N\}$ and away from the nodes of $\Psi$.

Then, locally, the Schrödinger equation

$$\mathrm{i}\hbar \frac{\partial}{\partial t}\Psi = \sum_{a=1}^{N} -\frac{\hbar^2}{2m_a}\Delta_a \Psi + V\Psi \tag{5.3}$$

with smooth potential $V$ is equivalent to the following system of PDEs:

$$m_a \left( \frac{\partial}{\partial t} + \sum_{b=1}^{N} (\vec{v}_b \cdot \nabla_b) \right) \vec{v}_a \equiv -\nabla_a V + \sum_{b=1}^{N} \frac{\hbar^2}{2m_b} \nabla_a \frac{\Delta_b \sqrt{\rho}}{\sqrt{\rho}} \tag{5.4a}$$

$$\frac{\partial \rho}{\partial t} + \sum_{a=1}^{N} \nabla_a \cdot (\rho\, \vec{v}_a) = 0 \tag{5.4b}$$

$$m_a \frac{\partial v_a^i}{\partial x_b^j} - m_b \frac{\partial v_b^j}{\partial x_a^i} \equiv 0. \tag{5.4c}$$

Those $N$-*body Madelung equations*, Eqs. (5.4), are dynamical equations for the time-dependent particle position probability density $\rho$ and the *(probability) current velocity (vector field)* $\vec{v} = (\vec{v}_1, \ldots, \vec{v}_N)$ (following Nelson's terminology [8]).

In particular, the continuity equation, Eq. (5.4b), assures probability conservation and states that the current velocity $\vec{v}$ governs the evolution of particle position probability for the ensemble (cf. Sec. 5.1 in [13]). Equation (5.4a) is naturally understood as a force equation for the $a$th body and Eq. (5.4c) constitutes a condition of vanishing 'quantum vorticity' [15]. The Madelung equations are also employed in de Broglie–Bohm theory [65–70] and stochastic mechanics [7,8,71–76].

There is an overt analogy to Newtonian continuum mechanics—more specifically the theory of Brownian motion. This was first outlined in detail by Fényes [7]. A formal analogy between the Schrödinger equation and the Fokker–Planck equation, as employed in the description of diffusion, was also noted by Schrödinger in 1931 [77] (see also [16,73,78–83]).

The analogy to Newtonian continuum mechanics provides us with a number of physically natural random variables. In this $N$-body theory they are the position of the $a$th body $\vec{r}_a$, its (current) momentum $\vec{p}_a = m_a \vec{v}_a$, its angular momentum $\vec{l}_a = (\vec{r}_a - \vec{r}_0) \times \vec{p}_a$ around a given $\vec{r}_0 \in \mathbb{R}^3$, and the total energy, as obtained from Eq. (5.4a)

$$E = \sum_{a=1}^{N} \frac{m_a}{2}\, \vec{v}_a^{\,2} + V - \sum_{a=1}^{N} \frac{\hbar^2}{2m_a} \frac{\Delta_a \sqrt{\rho}}{\sqrt{\rho}}\,. \tag{5.5}$$

In addition, one may introduce the *stochastic velocity (vector field)* for the $a$th body

$$\vec{u}_a = \frac{\hbar}{m_a} \mathrm{Re}(\nabla_a \Psi / \Psi) = \frac{\hbar}{2m_a} \nabla_a \rho / \rho. \tag{5.6}$$

The physical relevance of $\vec{u} = (\vec{u}_1, \ldots, \vec{u}_N)$ is justified on the grounds, that the so-called *quantum force* [65] on the right-hand side of Eq. (5.4a) may be reformulated in terms of it, and that the expectation value of the quantum-mechanical kinetic energy contains a summand of the type $\sum_{a=1}^{N} \frac{m_a}{2}\, \vec{u}_a^{\,2}$ (see for example [8,15,71,84,85]).

Notable derived quantities are the center of mass, the total momentum, and the total angular momentum (around any $\vec{r}_0$). How one defines kinetic energies ultimately depends on the particular interpretation of the stochastic velocity and we shall not commit to any particular one here.

Following this line of reasoning and putting the topic of the stochastic velocity aside for a moment, it may seem like a 'miracle' that the following holds: *The (Kolmogorovian) expectation value of any one of the aforementioned random variables is exactly equal to the expectation value of the corresponding quantum-mechanical operator.* For the 1-body theory, this was first observed by Takabayasi [86]. The author's contribution to this debate is the recognition that this constitutes *evidence for the validity of Kolmogorov's theory of probability in the quantum domain* (cf. [13] and Sec. 2.3 in [14]).

In this approach, ‘quantization’ still occurs: if one of the aforementioned random variables is a constant, then the induced probability measure is a Dirac measure. This is the case, for instance, for $E$ from Eq. (5.5), whenever $\Psi$ is a stationary state in the quantum-mechanical sense. As shown in Sect. 6, the notion of a stationary wave function may also be justified within this approach.

Of course, to obtain a mathematically rigorous theory, a restriction to local statements and requiring smoothness is not enough. Furthermore, if one ascribes such physical significance to the Madelung equations, then it is natural to build the theory thereon. In [13] this theory was named *geometric quantum theory*. To this end, R. and Poirier reviewed the mathematical theory of the Madelung equations in [15], concluding that the mathematical theory of the Madelung equations is still underdeveloped. While Gasser and Markowich did provide a suitable weak formulation [87], it is not entirely understood, how the Schrödinger equation can be reobtained [15]. As a side note, the work by Gasser and Markowich is of particular interest in this context, since the authors showed how to resolve the question of the “classical limit” in a natural manner (see [87], Rem. 3.1 in [15], and Sec. 5.1 in [13]).

Fortunately, for the purpose of constructing a mathematically rigorous quantum theory on the basis of Kolmogorov’s axioms, the contemporary lack of a mathematical theory of the Madelung equations does not pose a direct problem. Taking a pragmatic approach, we can carry over the dynamical evolution $t \mapsto U_t$ from quantum mechanics, and then only the corresponding physical random variables need to be rigorously defined. This may be understood as a *hybrid theory*, for it needs to rely on wave functions satisfying the Schrödinger equation as primary mathematical objects. This contrasts with the aim of geometric quantum theory as a research program, which is to erect the entire theory on Madelung’s equations and the Born rule in conjunction with Kolmogorov’s axioms [13].

## 6 A hybrid theory

For the purpose of formulating the hybrid theory (cf. Sect. 5), we constrain ourselves to (time-independent) Hamiltonians of the form

$$\hat{H} = \sum_{a=1}^{N} -\frac{\hbar^2}{2m_a}\Delta_a + V\,, \tag{6.1}$$

with $V$ real-valued and Lebesgue measurable. We further require that $\hat{H}$ is self-adjoint on some dense domain $\operatorname{dom}\hat{H}$ in $L^2(\mathbb{R}^{3N},\mathbb{C})$ and that $\operatorname{dom}\hat{H}$ is a subset of the Sobolev space $H^2(\mathbb{R}^{3N},\mathbb{C})$ (cf. Supplement 2 in [88]).

*Example 1* If $\hat{H}$ is given by Eq. (6.1) and $V$ is a Rellich class potential, then, by the Kato–Rellich theorem [89], $\hat{H}$ is self-adjoint on $\operatorname{dom}\hat{H} = H^2(\mathbb{R}^{3N},\mathbb{C})$. The most prominent class of examples are Coulomb Hamiltonians, which are of central importance in chemistry. ◇

The above restriction is made, because ‘observables’ always relate to particular models and it is the ‘observables’ relating to Eq. (5.3) (understood in a mathematically suitable sense) we consider here. Nonetheless, the main result in this context, Theorem 1 below, can in principle be generalized to other classes of models, though ultimately physical considerations dictate what random variables one ought to consider.

Among the random variables discussed in Sect. 5, only $\vec{v}_a$ and $E$ pose mathematical difficulties. $\vec{r}_a$ is well defined and $\vec{l}_a$ is defined, once $\vec{v}_a$ is. The following theorem solves this problem and shows that the aforementioned equality of expectation values also holds in this more general case.

**Theorem 1** *Let* $\hat{H}\colon \operatorname{dom}\hat{H} \to L^2(\mathbb{R}^{3N},\mathbb{C})$ *be given via Eq. (6.1) with real-valued and Lebesgue measurable* $V$ *such that* $\hat{H}$ *is self-adjoint on a dense domain* $\operatorname{dom}\hat{H} \subseteq H^2(\mathbb{R}^{3N},\mathbb{C})$ *in* $L^2(\mathbb{R}^{3N},\mathbb{C})$. *Furthermore, let* $\Psi_0 \in \operatorname{dom}\hat{H}$ *with* $\langle\Psi_0,\Psi_0\rangle = 1$. *For any* $t \in \mathbb{R}$, *set*

$$\Psi_t = \exp(-\mathrm{i}t\hat{H}/\hbar)\Psi_0. \tag{6.3a}$$

*Denote the corresponding probability measure by* $\mathbb{P}_t$ *(cf. Lemma 1 above).*

*Then, for all* $t \in \mathbb{R}$, *the following holds:*

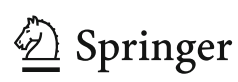

*(1)* $\Psi_t \in \operatorname{dom} \hat{H} \subseteq H^2(\mathbb{R}^{3N}, \mathbb{C})$ .

*(2) The following Radon–Nikodým derivatives exist, define $L^1$-random variables on $\big(\mathbb{R}^{3N}, \mathcal{B}^*(\mathbb{R}^{3N}), \mathbb{P}_t\big)$, and are given by the latter expressions $\mathbb{P}_t$-almost surely:*

$$(\vec{v}_a)_t = \frac{\frac{\hbar}{m_a} \operatorname{Im}\left(\Psi_t^* \nabla_a \Psi_t\right) \mathrm{d}^{3N} r}{\mathrm{d}\mathbb{P}_t} = \frac{\hbar}{m_a} \operatorname{Im}\left(\frac{\nabla_a \Psi_t}{\Psi_t}\right) \tag{6.3b}$$

$$E_t = \frac{\operatorname{Re}\big(\Psi_t^* \hat{H} \Psi_t\big) \mathrm{d}^{3N} r}{\mathrm{d}\mathbb{P}_t} = \frac{\operatorname{Re}\big(\Psi_t^* \hat{H} \Psi_t\big)}{|\Psi_t|^2}. \tag{6.3c}$$

*3) Their expectation values, $\mathbb{E}_t(\vec{v}_a)$ and $\mathbb{E}_t(E)$, satisfy the following identities:*

$$m_a\, \mathbb{E}_t(\vec{v}_a) = \langle \Psi_t, -\mathrm{i}\hbar \nabla_a \Psi_t \rangle \tag{6.3d}$$

$$\mathbb{E}_t(E) = \langle \Psi_t, \hat{H} \Psi_t \rangle. \tag{6.3e}$$

◇

The idea of using Radon–Nikodým derivatives to define $\vec{v}_a$ and $E$ is due to Gasser and Markowich [87]. A full proof is given in the appendix.

*Remark 2* (1) In full analogy to the definition of $\vec{v}_a$ in Theorem 1.(2), for $t \in \mathbb{R}$ we may introduce

$$(\vec{u}_a)_t = \frac{\frac{\hbar}{m_a} \operatorname{Re}\left(\Psi_t^* \nabla_a \Psi_t\right) \mathrm{d}^{3N} r}{\mathrm{d}\mathbb{P}_t} = \frac{\hbar}{m_a} \operatorname{Re}\left(\frac{\nabla_a \Psi_t}{\Psi_t}\right). \tag{6.4}$$

(2) The definition of the energy in Eq. (6.3c) is model-independent in the sense that it holds for any time-independent, densely defined, self-adjoint Hamiltonian $\hat{H}$ on $L^2(\mathbb{R}^{3N}, \mathbb{C})$ and wave function $\Psi : t \mapsto \Psi_t$ satisfying Eq. (6.3a) with $\Psi_0 \in \operatorname{dom} \hat{H}$. An analogous definition may be used for time-dependent Hamiltonians, though the treatment of domains requires more care (see for instance [90], Sec. V in [91], and [92]).

◇

The reader may have noticed that quantum mechanics makes weaker assumptions on the regularity of wave functions. However, for $L^2$-wave functions, the existence of the expectation value of, for instance, the quantum-mechanical energy is generally not guaranteed, so that even in quantum mechanics, some assumptions of weak differentiability are physically justifiable (cf. Refs. [15,85,93,94]). Still, it may be possible to generalize Theorem 1, so that the required order of differentiability on $\Psi_0$ is reduced by 1 (cf. Appx. A in [15]).

Similarly, the fact that $\mathbb{E}_t(\vec{r}_a)$ or $\mathbb{E}_t(\vec{l}_a)$ need not be defined for a given $\Psi_t$ does not pose any conceptual problems: they are defined if and only if the analogous quantum-mechanical expectation values are: the respective quantities are equal.

Before considering an explicit example, it is worth pointing out that both the time-dependent and the time-independent Schrödinger equation can be derived from the time evolution law in Eq. (6.3a). Thus, the theory predicts the same energies and stationary wave functions as quantum mechanics. Furthermore, in geometric quantum theory, the notion of a stationary wave function may be justified physically by imposing the requirement that the family of probability measures $(\mathbb{P}_t)_{t\in\mathbb{R}}$ is *stationary*. By definition, this holds, if $\mathbb{P}_t = \mathbb{P}_0$ for all $t \in \mathbb{R}$ (see also Def. 9.7.(v) in [33]). The following proposition provides a corresponding mathematical statement, which relies on known results used in the proof of Stone's theorem and is only stated for the convenience of the reader.

**Proposition 1** *Let $\hat{H}$, $\Psi_0$, and $(\mathbb{P}_t)_{t\in\mathbb{R}}$ satisfy the assumptions of Theorem 1 above.*

(1) *Then, for any $t \in \mathbb{R}$ there exists an element $\partial\Psi_t/\partial t$ in $L^2(\mathbb{R}^{3N}, \mathbb{C})$ such that*

$$\lim_{\varepsilon \to 0} \left\| \frac{\partial \Psi_t}{\partial t} - \frac{\Psi_{t+\varepsilon} - \Psi_t}{\varepsilon} \right\| = 0. \tag{6.5a}$$

*The map $\Psi : t \mapsto \Psi_t$ satisfies Eq. (5.3).*

(2) *If, in addition, the family of probability measures* $(\mathbb{P}_t)_{t\in\mathbb{R}}$ *is stationary, then there exists a constant* $\mathcal{E} \in \mathbb{R}$ *such that*

$$\hat{H}\Psi_0 = \mathcal{E}\Psi_0. \tag{6.5b}$$

*Moreover, E, as given via Eq. (6.3c), satisfies* $E_t = \mathcal{E}$ *for all* $t \in \mathbb{R}$.

◇

Again, the proof is given in the appendix.

*Example 2* We consider the simplest quantum-mechanical model of the hydrogen atom: $N = 1$ and $V(\vec{r}) = -\mathrm{e}^2/(4\pi\varepsilon_0|\vec{r}|)$. Using physics conventions, we denote the electron mass by $\mu$ and each stationary basis (wave) function by $\Psi_{nlm}$; $n$, $l$, and $m$ are the usual, appropriately chosen integers and $\mathcal{E}_n$ is the energy eigenvalue for the given $n$ (cf. Eq. (6.5b)).

We indeed have $\Psi_{nlm} \in \operatorname{dom}\hat{H} = H^2(\mathbb{R}^3, \mathbb{C})$. Recalling that $\Psi_{nlm}(t, \vec{r}) \equiv e^{-\mathrm{i}\mathcal{E}_n t/\hbar}\,\Psi_{nlm}(\vec{r})$, the respective random variables from Theorem 1.(2) are given via

$$\vec{v}_{(t,x,y,z)} = \frac{m\hbar}{\mu}\frac{1}{x^2+y^2}\begin{pmatrix}-y\\ x\\ 0\end{pmatrix} \tag{6.6a}$$

and

$$E_t(\vec{r}) = \mathcal{E}_n. \tag{6.6b}$$

The angular momentum around the origin, $\vec{l} = \vec{r} \times (\mu\vec{v})$, is not 'quantized', yet we may consider the angular momentum around the $z$-axis instead. Denoting the unit vector in $z$-direction by $\vec{e}_3$, the latter is given by

$$(\vec{r} - z\,\vec{e}_3) \times (\mu\vec{v}) = m\hbar\,\vec{e}_3. \tag{6.6c}$$

Note that, strictly speaking, the aforementioned random variables are only uniquely defined up to a set of measure zero and they are not defined for any $\vec{r}$ with $\Psi_{nlm}(\vec{r}) = 0$ (also a set of measure zero). Still, there is uniqueness in the sense that there is a maximal domain on which they are continuous; Eqs. (6.6) are to be understood in this manner. ◇

It is worth noting that the 'observables' $\vec{v}_a$ and $E$ from Theorem 1 depend on the particular ensemble one considers (as encoded by $\Psi_t$). This is in stark contrast to the momentum and energy operator in quantum mechanics, which generally only depend on the model, not the particular ensemble.

Finally, to physically compare geometric quantum theory and quantum mechanics, one needs to consider cases for which the theories make different predictions. The following example shows that such cases exist.

*Example 3* Continuing Example 3 above, we consider the initial wave function

$$\Psi_0 = a_1\Psi_{n_1 l_1 m_1} + a_2\Psi_{n_2 l_2 m_2} \tag{6.7a}$$

with $a_1, a_2 \in \mathbb{C}$ satisfying

$$|a_1|^2 + |a_2|^2 = 1. \tag{6.7b}$$

We determine the probability distribution for the energy in both theories, as given by Table 1.

In quantum mechanics and for $n_1 \neq n_2$, the probability to "measure" $\mathcal{E}_{n_1}$ is $|a_1|^2$, for $\mathcal{E}_{n_2}$, it is $|a_2|^2$, and for any other energy, it is zero. Hence, for nonzero $a_1$ and $a_2$, the distribution is supported on the set $\{\mathcal{E}_{n_1}, \mathcal{E}_{n_2}\}$ and it is independent of $t$.

In geometric quantum theory, on the other hand, for each $t \in \mathbb{R}$, the energy is given by the random variable

$$E_t = \frac{\mathcal{E}_{n_1}\left|a_1\Psi_{n_1 l_1 m_1}\right|^2 + \left(\frac{\mathcal{E}_{n_1}+\mathcal{E}_{n_2}}{2}\right) 2\mathrm{Re}\left(e^{-\mathrm{i}\omega t}\, a_1^*\Psi^*_{n_1 l_1 m_1} a_2\Psi_{n_2 l_2 m_2}\right) + \mathcal{E}_{n_2}\left|a_2\Psi_{n_2 l_2 m_2}\right|^2}{\left|a_1\Psi_{n_1 l_1 m_1}\right|^2 + 2\mathrm{Re}\left(e^{-\mathrm{i}\omega t}\, a_1^*\Psi^*_{n_1 l_1 m_1} a_2\Psi_{n_2 l_2 m_2}\right) + \left|a_2\Psi_{n_2 l_2 m_2}\right|^2} \tag{6.7c}$$

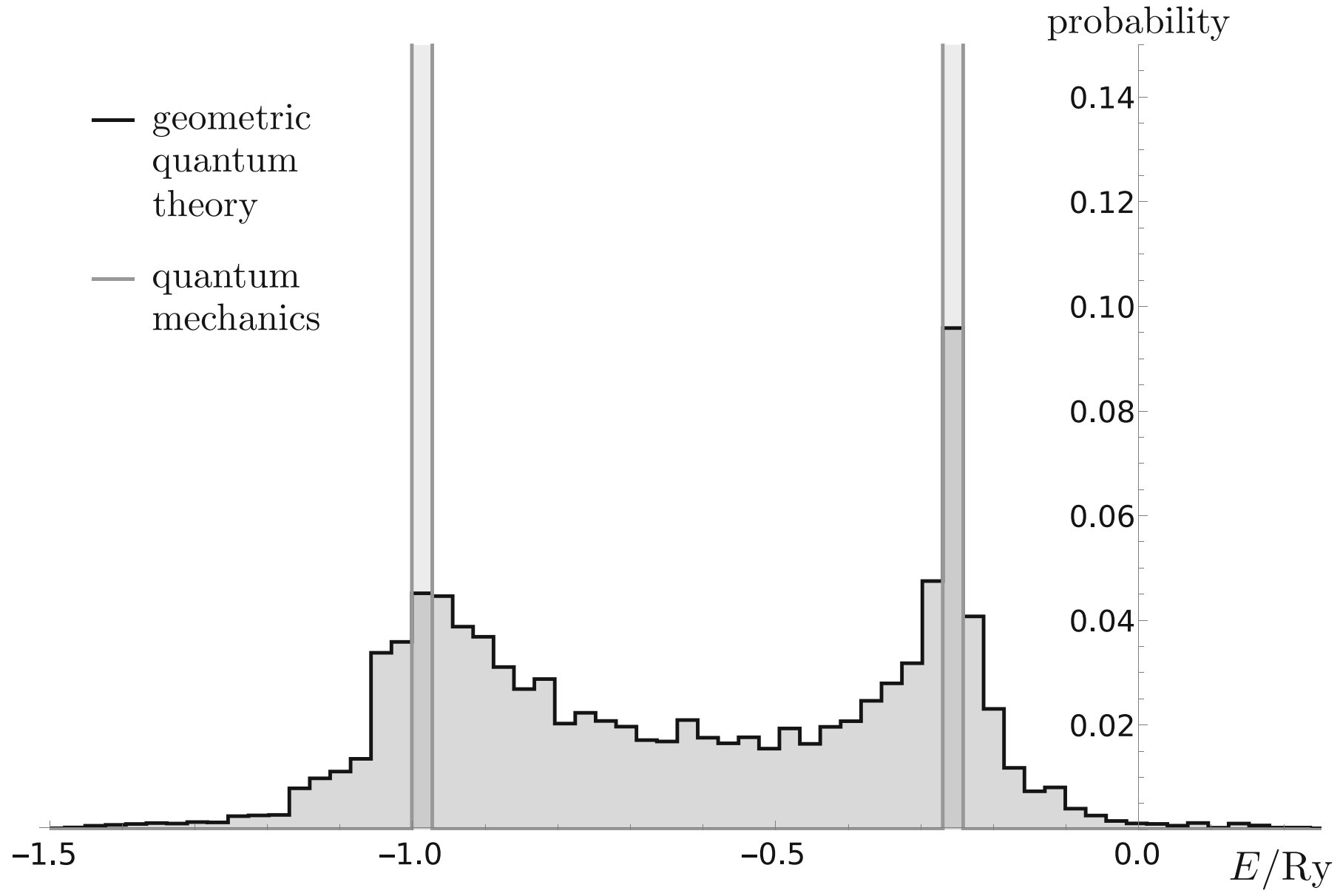

**Fig. 1** Numerically computed histograms comparing the energy probability distribution for the wave function in Eq. (6.7d), as predicted by geometric quantum theory and quantum mechanics. The quantum-mechanical distribution yields two "bins" of probability 0.5 each and is only indicated here

with $\omega = \left(\mathcal{E}_{n_2} - \mathcal{E}_{n_1}\right)/\hbar$. Accordingly, the distribution is $\mathbb{P}_t \circ (E_t)^{-1}$. If $n_1 = n_2$ or if either $a_1$ or $a_2$ are zero, the quantum-mechanical prediction is reproduced. Yet otherwise, there is a qualitative difference, for energies other than $\mathcal{E}_{n_1}$ and $\mathcal{E}_{n_2}$ may be attained.

Figure 1 depicts numerically computed histograms of the two distributions for the initial state

$$\Psi_0 = \frac{1}{\sqrt{2}}\,\Psi_{100} + \frac{\mathrm{i}}{\sqrt{2}}\,\Psi_{211}. \tag{6.7d}$$

Both histograms exhibit peaks around the values $\mathcal{E}_1$ and $\mathcal{E}_2$, yet they are less sharp in geometric quantum theory. It may seem counterintuitive that geometric quantum theory allows for energies which are smaller than $\mathcal{E}_1$ and larger than $\mathcal{E}_2$ in this case, yet this prediction is physically justifiable: $\Psi_{100}$, $\Psi_{211}$ and $\Psi_0$ describe physically distinct ensembles and physical intuition derived from the quantum-mechanical concept of "superposition" need not apply here.

It is worth noting that the numerical simulations for this prediction are consistent with $t \mapsto \mathbb{P}_t \circ (E_t)^{-1}$ being stationary. There may exist a mathematical proof thereof, plausibly even for more general "bound states". ◇

It is important to note that it is not possible to put the differing predictions in Example 3 above to experimental scrutiny, for the model describes an *isolated* hydrogen atom (a "bound state"). Generally speaking and as discussed in more detail in part III of this series, to predict the outcomes of "measurements" other particles or radiation probing the system need to be included in the model (see Sec. 5.7 in [95] for an example). The model of Example 2 may be used for making predictions in this sense, if one employs the electron to 'probe' the proton instead. That is, one may be able to obtain differing empirical predictions by considering time-dependent electron–proton scattering instead.

## 7 Conclusion

It was shown how to construct a mathematically rigorous, non-relativistic quantum theory for an $N$-body system subject to a time-independent scalar potential, which accords with Kolmogorov's theory of probability. The two main steps were, first, to employ the Born rule for position to construct the probability space (Lemma 1), and, second, to define physically appropriate random variables in relation to the dynamical theory borrowed from quantum mechanics (Theorem 1).

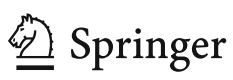

*This theory consists of a (time-dependent) probability space, a collection of (time-dependent) random variables, as well as a time evolution law for those two. Mathematical probability theory then dictates how to compute probabilities, probability distributions, expectation values, etc.*

In this sense, the reader is provided with the following synopsis of the theory:

*Time evolution*: The Hamiltonian $\hat{H}$ satisfies the assumptions of Theorem 1 and the initial wave function $\Psi_0$ is required to be in dom $\hat{H}$. For any $t \in \mathbb{R}$, $\Psi_t$ is then given via Eq. (6.3a).
*Probability space*: At time $t \in \mathbb{R}$ the probability space is $\big(\mathbb{R}^{3N}, \mathcal{B}^*(\mathbb{R}^{3N}), \mathbb{P}_t\big)$, as given by Lemma 1. In particular, the probability measure $\mathbb{P}_t$ for the positions of the $N$ bodies in $\mathbb{R}^{3N}$ is defined via the probability density $\rho_t = |\Psi_t|^2$ (cf. Eq. (4.1)).
*Random variables*: The central, physical random variables are the position of the $a$th body $\vec{r}_a$, its (current) velocity $\vec{v}_a$ as defined in Eq. (6.3b), its stochastic velocity as defined via Eq. (6.4), and the total energy as defined via Eq. (6.3c). Notable random variables, which are derived from the aforementioned ones, are the various angular momenta, the center of mass, and the total momentum (cf. Sect. 5).

The requirement that $\Psi_0 \in$ dom $\hat{H}$ constitutes a stronger regularity condition than imposed in the analog quantum-mechanical theory. Furthermore, no physical significance is ascribed to the respective quantum-mechanical operators, which is independent from that of the aforementioned random variables, though they may still be of use in calculating expectation values, etc.

This is a "hybrid theory", because, philosophically, geometric quantum theory views the time evolution law above as a consequence of the respective Madelung equations. Accordingly and contrary to what was done in this work, the wave functions $\Psi$ are (indirectly) defined in terms of $\rho$ and $\vec{v}$ (cf. [15]).

The presented approach of constructing such a hybrid theory can in principle be applied to other quantum-mechanical models, though one does need additional physical arguments to motivate the random variables of physical interest.

To ease further research in this direction, a brief guideline for how to do so is given as follows:

(1) Consider the quantum-mechanical analog model in the "position representation". Define the (time-dependent) probability space via the respective Born rule for position, as done in Sect. 4 and Lemma 1, in particular.
(2) For the given Hamiltonian find a (heuristic) Madelung-type reformulation of the respective Schrödinger equation. Geometric quantum theory asserts that for physically acceptable models, such a reformulation always exists (cf. Secs. 3 and 4 in [13] and Sec. 2.3 in [14]). This Schrödinger equation will also hold in geometric quantum theory, with the difference that the (initial) wave functions must be elements of the domain of the Hamiltonian (cf. Theorem 1.(1)).
(3) As in Sect. 5, employ the analogy to Newtonian (continuum) mechanics to motivate the physically relevant random variables. Those are to replace the respective quantum-mechanical observables. As in Theorem 1, Radon–Nikodým derivatives may be employed to define those random variables in a mathematically rigorous manner (cf. Theorem 1.(2)) and to relate their expectation values to the quantum-mechanical ones (cf. Theorem 1.3)).

With regards to the foundations of mathematical probability theory, it is noteworthy that the equality of the quantum-mechanical and the physically natural 'Kolmogorovian' expectation values may be interpreted as evidence for the validity of Kolmogorov's theory in the quantum domain. This was already shown heuristically in [13] and Chap. 2 of [14], yet this work provides a mathematically rigorous argument.

Of course, it is ultimately an empirical question whether this approach provides a physically viable competitor to quantum mechanics. In this respect, differences in prediction are to be explored.

As illustrated by Example 3, the distributions and higher order moments of observables other than position generally differ between the two theories.

A physical problem encountered in Example 3 is that testing such systems may go beyond the model employed, because the radiation, for instance, needs to be included in the dynamics. In turn, to make specific experimental predictions, models accounting for spin and time-dependent electromagnetic radiation may need to be considered. Madelung-type reformulations of the Pauli equation are indeed known [96,97], so that a consideration of the corresponding random variables may provide clues to an experiment.

Finally, the example of angular momentum in Example 2 suggests further conceptual differences, some of which may enable more direct experimental inquiries.

**Acknowledgements** R. would like to thank Bill Poirier and Mike Scherfner for their indirect support of his research efforts. He also thanks Jamal Berakdar for helpful comments, Wolfgang Paul for helpful discussion on the BCHSH inequality, and Mario Flory for inviting him to a conference, where he held a similarly titled talk. Furthermore, R. is indebted to Anhalt UAS for enabling his participation therein.

**Funding** Open Access funding enabled and organized by Projekt DEAL. Open access funding enabled and organized by Project DEAL and Anhalt University of Applied Sciences.

**Data availability** Computational source code for the data used in Fig. 1 is available on the website of the journal as supplementary material to this article.

**Declarations**

**Conflict of interest** The author does not have any conflict of interest to declare.

## Appendix

*Proof of Theorem 1* (1) The assertion follows directly from the invariance of dom $\hat{H}$ under time evolution (see for instance Prop. A.1 in [15]).

(2) The two (signed, vector-valued) measures in the "numerator" are well defined and finite, since $\Psi_t \in H^2(\mathbb{R}^{3N}, \mathbb{C})$. Now, to apply the Radon–Nikodým theorem (cf. Cor. 7.34 and Exercise 7.5.1 in [33]), we must show that for all $W \in \mathcal{B}^*(\mathbb{R}^{3N})$ with $\mathbb{P}_t(W) = 0$, either measure of $W$ at time $t$ also vanishes. For this purpose, take any representative $\Psi_t$ of the corresponding $L^2$-equivalence class. Then, $\mathbb{P}_t(W) = 0$ implies that $|\Psi_t|^2 = 0$ almost everywhere on $W$ (with respect to the Lebesgue measure; cf. Thm. 4.8 in [33]). Thus, $\Psi_t = 0$ almost everywhere on $W$. The respective, latter expressions for $(\vec{v}_a)_t$ and $E_t$ follow from direct computation.

(3) This holds by construction. □

*Proof of Proposition 1* (1) For the first statement, see Thm. VIII.7 in [32]. This also implies that Eq. (5.3) holds. The right-hand side of Eq. (5.3) is well defined because of Theorem 1.(1).

(2) By Thm. 4.8.(i) in [33], we have $|\Psi_t|^2 \equiv |\Psi_0|^2$ in $L^1$. Upon taking the square root, we find that there exists a real-valued function $\alpha\colon t \mapsto \alpha(t)$ such that $\Psi_t \equiv e^{-\mathrm{i}\alpha(t)/\hbar}\Psi_0$ in $L^2$. Now apply the spectral theorem (Thm. VIII.4 in [32]) to rewrite $\hat{H} = U^{-1}\hat{\mathcal{E}}U$ for some multiplication operator $\hat{\mathcal{E}}$ on the $L^2$-space of some measure space $(\Omega, \mathcal{A}, \mu)$ and for some unitary $U$. From the definition of $\exp(-\mathrm{i}t\hat{H}/\hbar)$, it follows that:

$$\left(\exp(-\mathrm{i}t\hat{\mathcal{E}}/\hbar) - \exp(-\mathrm{i}\alpha(t)/\hbar)\right)(U\Psi_0) \tag{A.1}$$

vanishes in $L^2$ for all $t \in \mathbb{R}$. Now, pick a representative of $U\Psi_0$. Then, there exists a set of full measure $\Omega'$, such that for the representative $U\Psi_0$, the above expression vanishes pointwise on that set. Since $U\Psi_0 \neq 0$, there exists an $\omega_0 \in \Omega'$ with $(U\Psi_0)(\omega_0) \neq 0$. Writing $\big(\hat{\mathcal{E}}U\Psi_0\big)(\omega_0) = \mathcal{E}(\omega_0)\,(U\Psi_0)\,(\omega_0)$, it follows that $\mathcal{E}(\omega_0)\,t \equiv \alpha(t)$. This shows the first assertion. The second one holds by definition of $E$. □